\documentclass[aip,apl,amsmath,amssymb,reprint,longbibliography]{revtex4-1}

\usepackage{graphicx}
\usepackage{dcolumn}
\usepackage{bm}
\usepackage{float}
\usepackage[utf8]{inputenc}
\usepackage[T1]{fontenc}
\usepackage{mathptmx}
\usepackage{etoolbox}
\usepackage{xcolor}
\usepackage{gensymb}
\makeatother
\begin{document}

\preprint{AIP/123-QED}%

\title[]{Cesium Based Laser-Atomic Oscillator}
\author{Saurabh Pandey}
\affiliation{Sandia National Laboratories, Albuquerque, New Mexico 87123, USA}
\author{Roger Ding}
\affiliation{Sandia National Laboratories, Albuquerque, New Mexico 87123, USA}
\author{George Burns}
\affiliation{Sandia National Laboratories, Albuquerque, New Mexico 87123, USA}
\author{Yuan-Yu Jau}
\email{yjau@sandia.gov}
\affiliation{Sandia National Laboratories, Albuquerque, New Mexico 87123, USA}


\date{\today}

\begin{abstract}
We report the first demonstration of a laser-atomic oscillator with cesium (Cs) atoms.
A laser-atomic oscillator (LAO) is analogous to an active mode-locked laser with a self-excited modulator, i.e. atoms, at a ground-state hyperfine transition frequency.
Therefore, a LAO can be configured as the simplest active atomic clock or a self-oscillating, earth-field atomic magnetometer that delivers oscillation signals both optically and electrically.
With the current experimental Cs-LAO setup, when it is configured as an atomic clock using the 0--0 hyperfine transition, the short-term fractional frequency instability is around 10$^{-10}$ level.
When it is configured as a self-oscillating magnetometer using a magnetically-sensitive hyperfine transition, the magnetic field sensitivity is around 100 fT/$\sqrt{\rm{Hz}}$ at 60 Hz.
The presented Cs-LAO uses a cavity length from $\sim6.5$ cm to $\sim11.4$ cm.
Ultimately, the minimal length of a Cs-LAO device can be $\leq1.63$ cm.
Our new efforts unlock the potential of building truly chip-scale atomic clocks and magnetometers.

\end{abstract}

\maketitle

Accurate and high-precision timing/frequency devices are an essential need in many areas today, e.g., fundamental timekeeping, high-speed data transfer, global positioning system (GPS) navigation, cell phone towers, secure communication, etc., and atomic clocks have long served the role of primary timekeeping \cite{Lombardi2007JMS}.
A miniaturized atomic clock with low size, weight, and power (SWaP) holds value in standalone precision timing and frequency referencing without relying on an external calibration signal, thereby enabling applications such as GPS-denied navigation, anti-jamming communication, high-performance digital electronics, etc.
For these reasons, a highly miniaturized Chip-Scale Atomic Clock (CSAC) \cite{Lutwak2007} was first commercialized in 2011, and these CSACs have been developed over the past two decades.
Most atomic frequency standards are based on the passive atomic clock technique where an external local oscillator is stabilized to an atomic reference.
For a CSAC that utilizes coherent population trapping \cite{Vanier2005APB}, the inevitable peripheral microwave local oscillator and control electronics make further miniaturization very challenging.
Thus, the demonstrated CSACs \cite{Travagnin2021} have a volume $\geq$ 15 cm$^3$.
Therefore, the desire to pursue new atomic-clock devices with smaller SWaP and better timing precision remains.

In contrast to the complex electronics in a passive atomic clock, an active atomic clock can be fundamentally simpler because the atoms emit the clock signal directly, eliminating the need for a separate local oscillator and associated control electronics.
Well-known active atomic clocks, such as ammonia \cite{Gordon1955PR}, hydrogen \cite{Goldenberg1960PRL}, and rubidium masers \cite{Vanier1968PR, Davidovits1964APL}, however, have been limited in size by their microwave cavities that must match the atomic-clock transition frequency.
To eliminate the microwave cavity, the simple, low-power laser-atomic oscillator (LAO) was proposed as an active atomic clock enabled by combining a laser gain element, an intra-cavity alkali-vapor cell, and push-pull optical pumping \cite{Jau2004PRL}.
For a LAO, only one of the device dimensions has to match the microwave wavelength, greatly reducing the device volume.
The first proof-of-concept experiment for the LAO was published in 2007 using $^{39}$K (potassium) vapor \cite{Jau2007PRL,Jau2008IEEE}. 
This $^{39}$K-LAO generated clock signals as optical and electrical pulse trains with a repetition rate at the $^{39}$K ground-state hyperfine splitting frequency at 461.7 MHz.
However, this relatively low frequency makes the $^{39}$K-LAO vulnerable to the magnetic field fluctuation and is therefore not ideal for making a stable atomic clock.
Its corresponding hyperfine transition wavelength at 32.5 cm also prevents the $^{39}$K-LAO from miniaturization. 

To advance the LAO technology toward a truly chip-scale atomic clock, we demonstrate a LAO based on $^{133}$Cs.
The cesium hyperfine frequency, 9.2 GHz, naturally enables a $^{133}$Cs-LAO to be smaller ($\leq0.1$ cm$^3$) because the fundamental first-order cavity length can be $\leq1.63$ cm.
Due to the larger ground-state hyperfine frequency in cesium than in potassium, the device is also less sensitive to changes in the magnetic field.
Lastly, this same device can also be operated as a highly miniaturized and self-oscillating magnetometer by simply tuning the external cavity length to lock it to a magnetically sensitive Zeeman sublevel pair.

In this Letter, we present results on the realization of a $^{133}$Cs-LAO.
We discuss results based on fourth-order ($\sim6.5$\,cm) and seventh-order ($\sim11.4$\,cm) cavities.
The 9.2\,GHz clock signal is a low phase noise, narrow linewidth signal that has a short-term fractional frequency instability of the order of $10^{-10}$.
The clock signal drifts on a timescale of tens of seconds, mainly due to changes in the effective cavity length due to thermal drift in the external cavity components and to the temperature sensitivity of the vapor cell buffer-gas mixture.

\begin{figure}
    \centering
    \includegraphics[width=1\columnwidth]{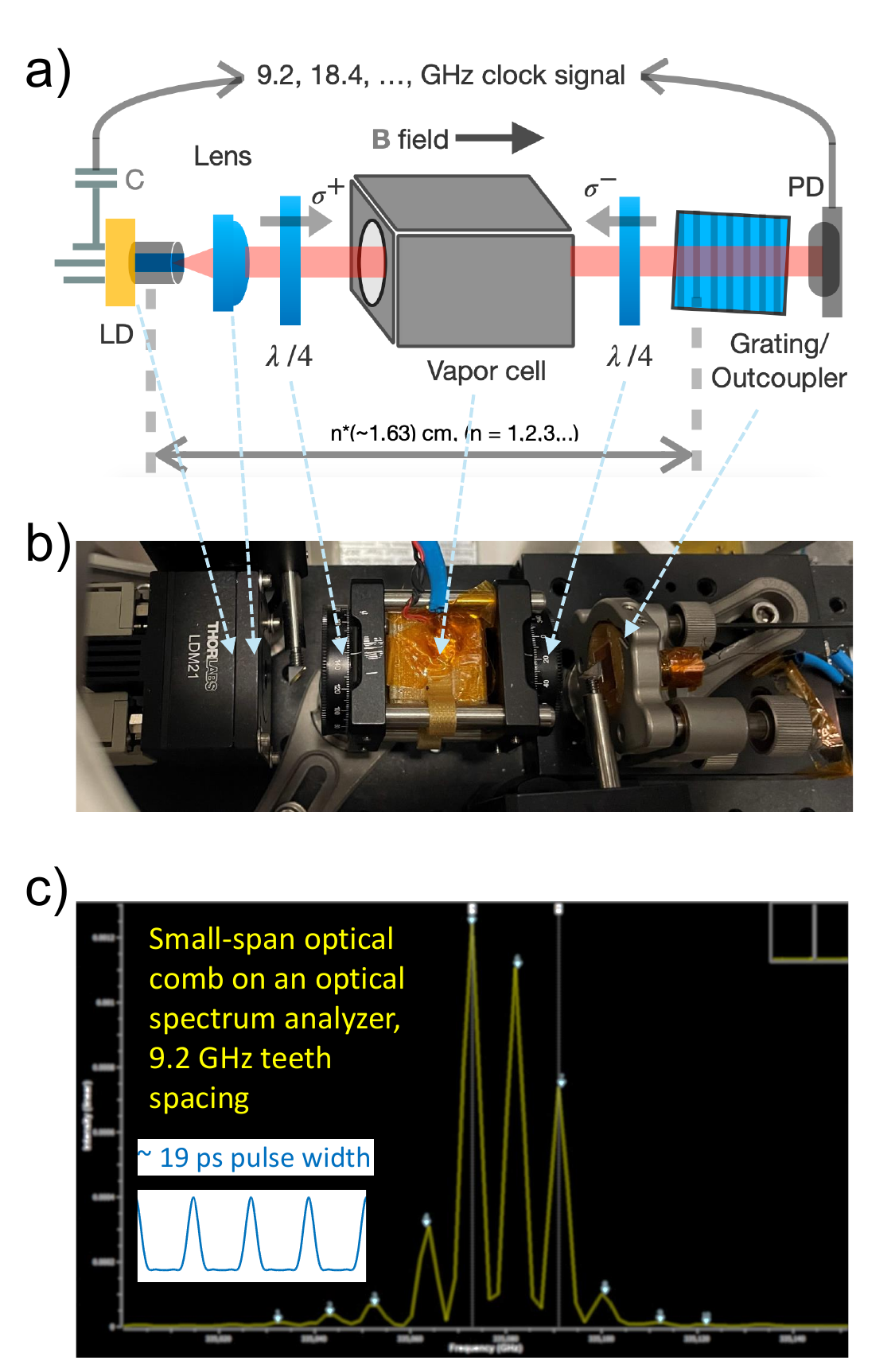}
    \caption{
        \label{fig:figure1}
        a) Scheme of the laser atomic oscillator. LD, PD, C stands for laser diode, photodiode, and capacitor, respectively.
        b) A fourth order Cs-LAO setup.
        c) Optical spectrum of 0-field LAO signal containing the carrier at 895nm and multiple sidebands 9.2\,GHz apart.
        The y-axis is on a linear scale.
        Each block on the x-axis is 5 GHz and y-axis starts with 0.
        The span of the optical teeth is consistent with the optimal PPOP spectrum calculated in reference \cite{Post2005PRA}.}
\end{figure}

The mechanism of a LAO relies on push–pull optical pumping (PPOP) of the atoms with modulated light where alternating circular polarization drives the atoms into a coherent superposition of the two ground hyperfine sublevels, building population in the clock states or a magnetically sensitive pair \cite{Jau2004PRL}.
This changes the atomic absorption (or loss) in the cavity, which in turn feeds back onto the laser, sustaining modulation — coupling atomic coherence, atomic absorption, and laser cavity dynamics into a self-consistent oscillator \cite{Jau2007PRL}.
Because the atoms modulate the optical loss inside the laser cavity at the hyperfine frequency and this modulation feeds back to the laser light (and its diode bias), one obtains an output that is a stable microwave signal without needing a separate microwave source. 
The modulated light (or the modulated electrical impedance across the diode) can be used to recover the microwave signal at the atomic transition frequency.
Essentially, the atomic transition directly controls and stabilizes the laser’s temporal behavior, turning passive atoms and an active laser with cavity feedback into a self-locked oscillator at the atomic clock frequency.

We now describe the key elements of the LAO.
As shown in Fig.\ref{fig:figure1}, the LAO is an external cavity consisting of an atomic vapor cell, two quarter waveplates, and a grating.
The light source is a low power, broadband laser gain chip with very low front-facet reflectivity (< 10$^{-5}$) and gain center near 895\,nm ($D_1$ line of Cs).
We use Innolume gain chip (GC-920-90-TO-200-B) with a cavity length of 1.5\,mm, or, internal mode spacing of about 36\,GHz.
It is better to have large internal mode spacing (shorter gain chip cavity length) for less interference with the microwave LAO signal. 
A shorter gain chip also contributes less to the effective cavity length instability.
The volume Bragg grating (VBG) at the other end of the external cavity is from OptiGrate.
We worked with gratings from Ondax (Coherent) as well and got similar LAO results.
The grating has a full width at half maximum (FWHM) bandwidth of 0.07\,nm.
It is important to match the grating bandwidth with that of the atomic spectral width.
The grating is written at a 5\,deg angle to avoid the front surface reflection from coupling back to the gain chip and being amplified.
Avoiding optical feedback holds true for other optic interfaces as well, e.g., cell windows and detection optics after the VBG.
The vapor cells from SRI International had cesium and nitrogen ($\text{N}_2$) buffer gas filling at a pressure of 66.6\,psi (3444 Torr).
Due to power loss through the heated vapor cell (up to 25\% in single pass), we make the input beam slightly elliptic in polarization and then compensate the power loss by using < $\lambda/4$ retardance in the other half of the cavity.
It is important to minimize the intra-cavity loss with anti-reflection coating at the optical interfaces.
Finally, not shown in Fig.1, there are three pairs of Helmholtz coils to either null the magnetic field (B-field) at the vapor cell location or to apply a finite field along the cavity long axis.

We first characterize the laser gain chip using an optical spectrum analyzer (Bristol 771) to confirm it does not lase by itself up to the maximum allowed current, the internal mode strength is negligible, and the mode spacing is consistent with the cavity length.
The grating bandwidth is characterized using the same gain chip where light transmitted through the grating is routed to an optical spectrum analyzer and the resulting intensity profile corresponds to the grating bandwidth.
The vapor cell spectral width and center is characterized by measuring the PPOP signal strength of the 0-0 transition with respect to a seed beam frequency (see below).
For the PPOP measurements, a DBR laser (Vescent D2-200 at 895\,nm) is used to generate alternating circular polarization pulses ($\sigma^{+}$ $\leftrightarrow$ $\sigma^{-}$) at 9.2\, GHz by the use of an intensity modulator and a Michelson interferometer \cite{Liu2013PRA}.
By scanning this seed laser wavelength, we measure a FWHM Lorentzian linewidth of the atomic spectrum to be about 40\,GHz.
We study the PPOP contrast with respect to the temperature of the vapor cell  and typically operate at $\sim$ 90\,\degree C where the 0-0 PPOP contrast is slightly above 1$\%$.

The external cavity is first aligned to lase without the cell and roughly at the center of the atomic spectrum (at about 335080\,GHz). 
This is done by tuning the grating temperature, stabilizing at about 45\,\degree C.
We then insert the cell and heat it to 90\,\degree C.
The cavity length is further fine tuned by monitoring the LAO cavity modes on a fast photodiode (Thorlabs RXM42AF) and the mode frequency is brought close to 9.2\,GHz.
We get a precise microwave frequency reference from the vapor cell characterization by centering the 0-0 signal in a narrow microwave frequency sweep.
In order to make the initial alignment more efficient, we shine the seed beam through the cell at an angle to overlap with the intra-cavity beam at the cell.
With the help of the PPOP signal encoded on this seed beam, we fine tune the B field to near zero (collapse all seven $\Delta m_f=0$  lines to the degenerate signal, \textit{F} = 3 to \textit{F} = 4).
We call this a 0-field signal, which is stronger than any individual Zeeman sublevel pair signal.
The intra-cavity light, exiting from the grating, starts to contain a weak LAO signal induced by the seed.
This helps in coarse tuning for e.g., cavity length, waveplate angles, gain chip current, etc. 
Finally, to get to 0-0 or other lines, we increase the B field to a finite value (typically < 1\,G). 

\begin{figure}
    \centering
    \includegraphics[width=1\columnwidth]{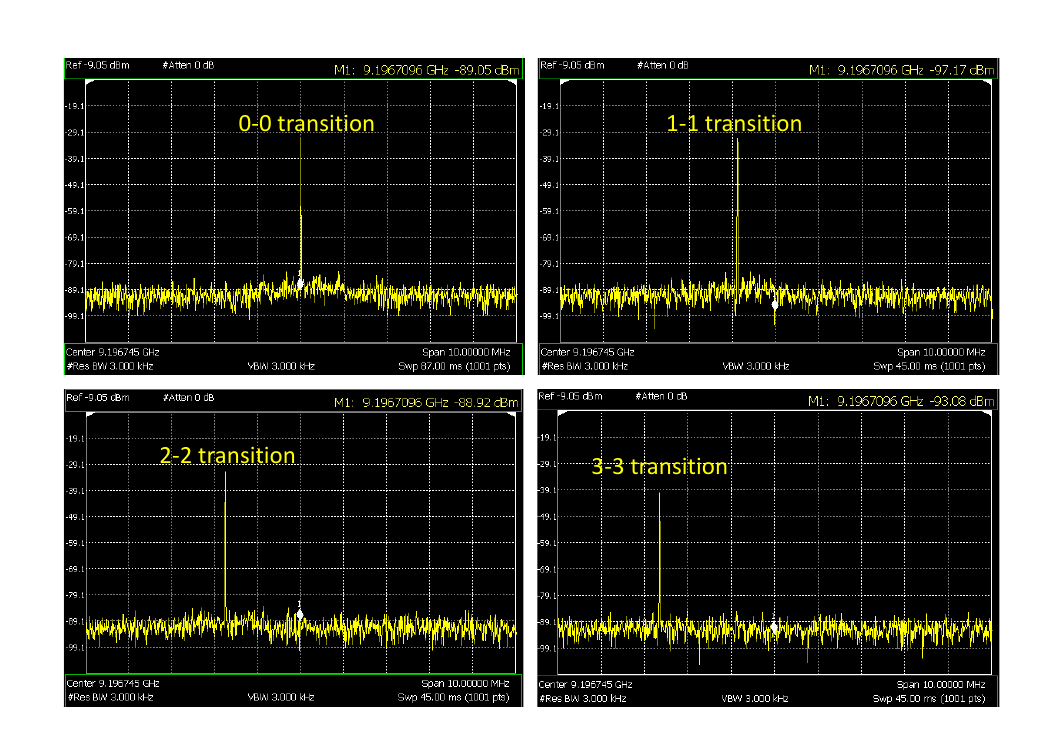}
    \caption{
        \label{fig:figure2}
        The LAO signal locked to different magnetic sublevel pairs ($\Delta m_f=0$) when a 1.3~G magnetic field is applied.}
\end{figure}

\begin{figure}
    \centering
    \includegraphics[width=1\columnwidth]{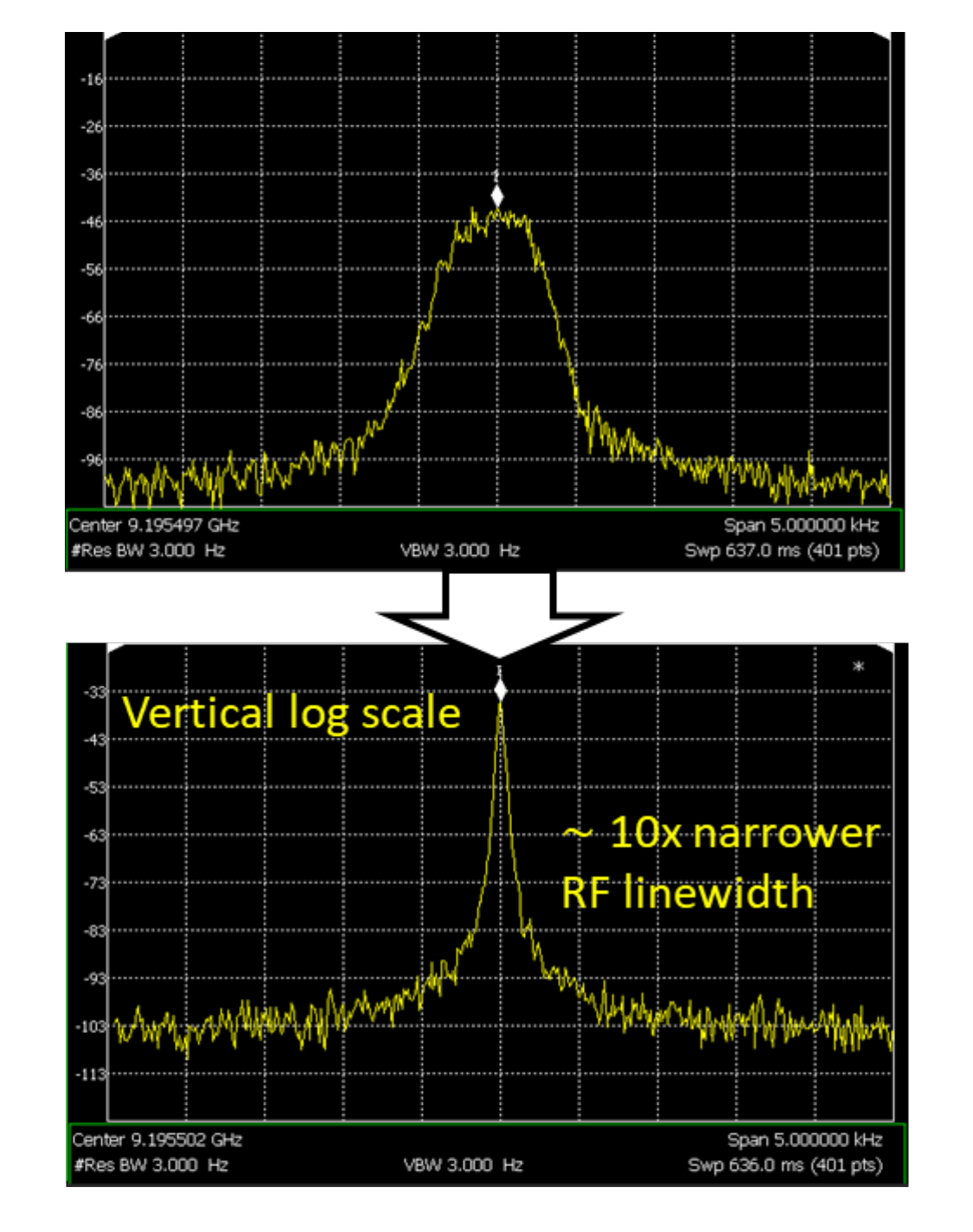}
    \caption{
        \label{fig:figure3}
        Improving the phase noise in the 0-0 signal by minimizing unwanted external optical feedback and enclosing the setup.}
\end{figure}

\begin{figure}
    \centering
    \includegraphics[width=1\columnwidth]{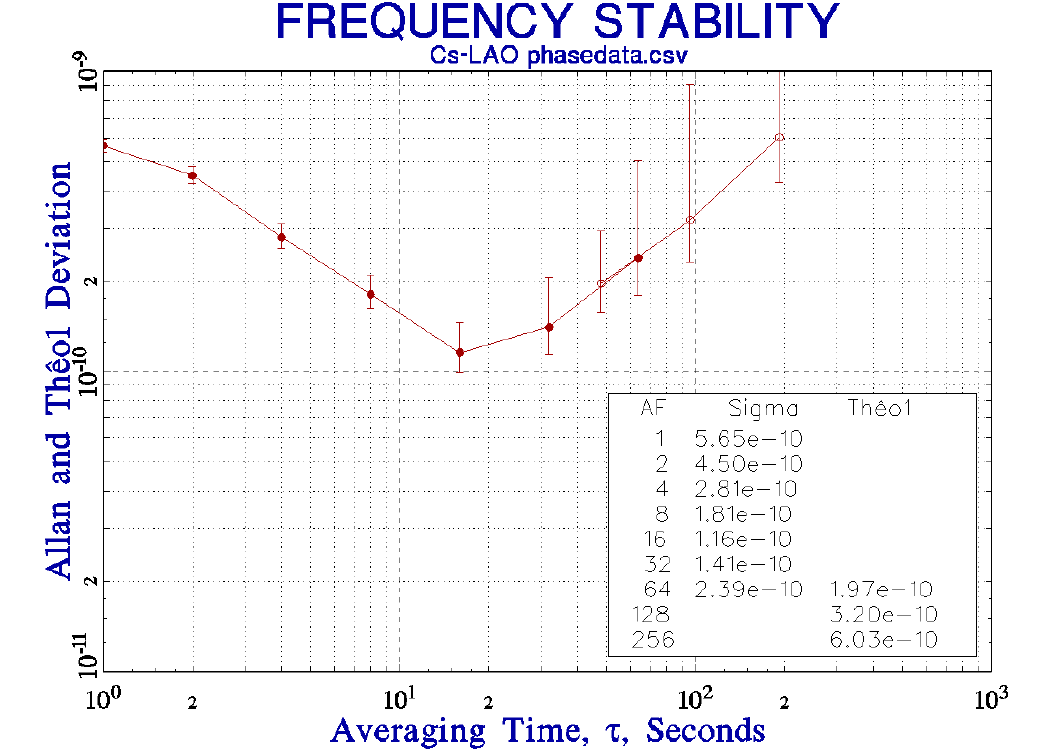}
    \caption{
        \label{fig:figure4}
        Short-term Allan deviation of the LAO locked to the 0-0 hyperfine clock transition.}
\end{figure}

\begin{figure}
    \centering
    \includegraphics[width=1\columnwidth]{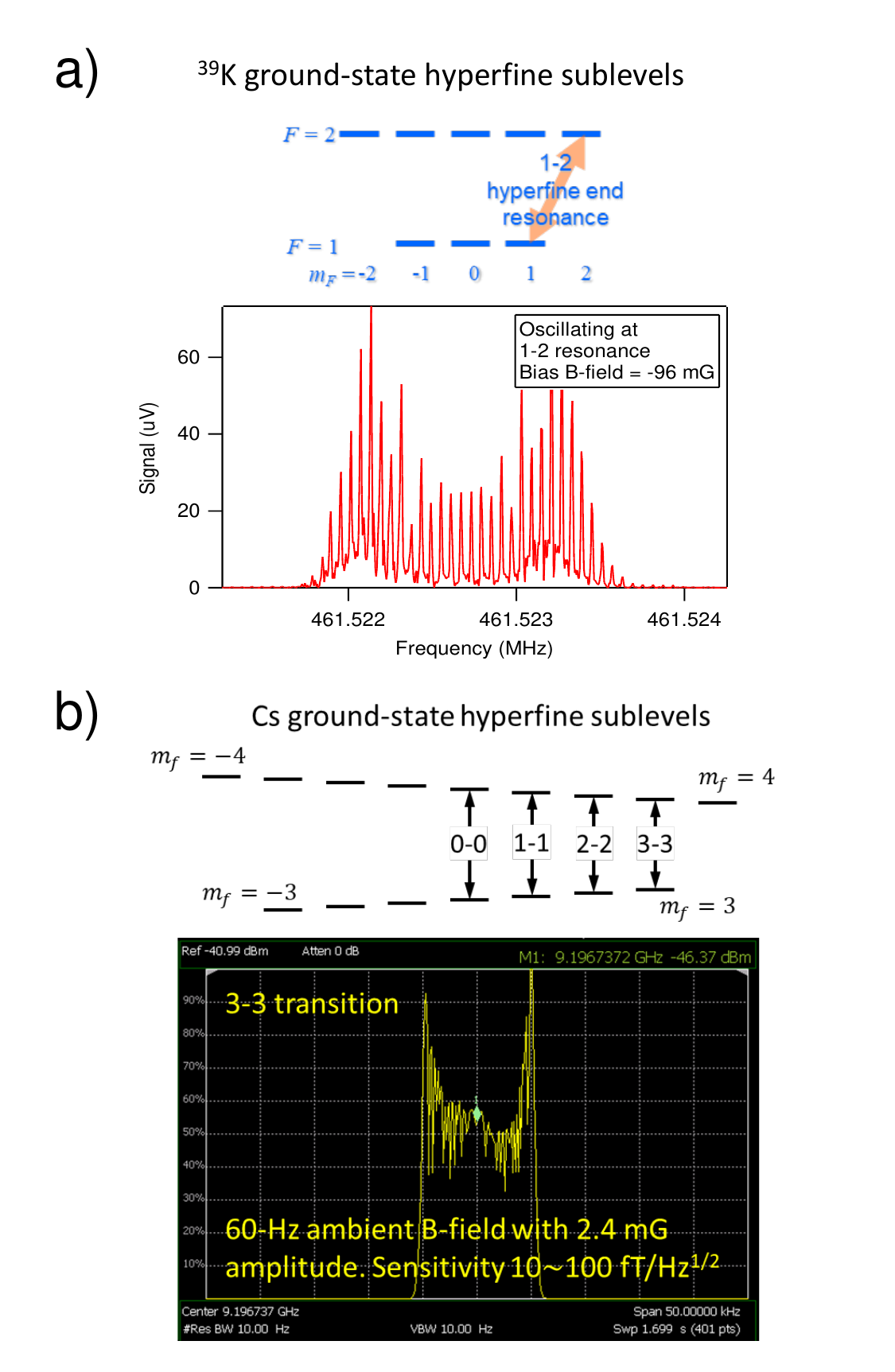}
    \caption{
        \label{fig:figure5}
        B-field modulated oscillations for, a) potassium \cite{Jau2009PU}, and b) cesium.}
\end{figure}

To get a LAO signal from a given Zeeman sublevel pair with $\Delta m_f=0$, a number of knobs can be used- gain chip current, grating angle, and cavity length.
Fig.\ref{fig:figure2} shows 0-0 to 3-3 transition signals with a 1.3\,G magnetic field (see Fig.\ref{fig:figure5}b for the Cs hyperfine sublevels of interest in this work).
After optimizing the cavity length and gain chip current with respect to the signal strength and stability, we use the grating angle to do the fine tuning and switching to a different Zeeman sublevel pair.
Fig.\ref{fig:figure3} shows zoom-in on the 0-0 signal.
With better temperature stability of the external cavity elements (gain chip, cell, and grating), vibration reduction, and enclosing the setup, we were able to narrow the signal linewidth by about an order of magnitude.
Note that the horizontal scale is 500\,Hz/div and log scale along the vertical axis.
The FWHM linewidth of the 0-0 signal is at the few Hertz level.
In Fig.\ref{fig:figure4}, we plot the computed Allan deviation of the 0-0 signal.
We believe the turn around of the averaged frequency signal at about 20\,s is mainly due to the cell temperature instability.
The cuboid-shaped cell, 0.5" x 0.5" x 0.24", is heated from the sides only while the larger surface area front and back facets are exposed to air.
We studied the contribution of other factors to the clock signal stability as well, such as the gain chip current (at $\sim$ 10Hz/$\mu$A), grating temperature (at $\sim$ 10Hz/mK), and cavity length (cavity pulling coefficient of $\sim 10^{-4})$, all with the nonideal buffer gas mixture. 
These challenges can be circumvented in a more compact first-order-cavity setup and using precise $\text{N}_2$-Ar buffer gas mixture to make the atomic resonance dependence on the temperature only 2$^\text{nd}$-order sensitive \cite{Vanier1982JAP, Kozlova2011PRA}.

To demonstrate operation of the same device as a high-sensitivity atomic magnetometer, we lock the LAO to the 3-3 transition, see Fig.\ref{fig:figure5}b.
Among all the $\Delta m_f=0$ pairs, 3-3 gives the highest B-field sensitivity.
Note that switching to a different transition can be done via the gain chip current or the cavity length.
We observe broadening of the B-field sensitive pair signals due to the mains 60\,Hz.
For the data in Fig.\ref{fig:figure5}b, we deduce a sensitivity of 10--100 fT/$\sqrt{\text{Hz}}$. 
Another example in Fig.\ref{fig:figure5}a is when the $^{39}$K-LAO was configured to operate on the magnetic-field-sensitive 1–2 hyperfine end transition, its output frequency becomes strongly modulated by the ambient 60-Hz field, producing a characteristic FM spectrum\cite{Jau2009PU}. 
In summary, this shows that a chip-scale LAO can also serve as an Earth-field, ultra-low-SWaP, high-sensitivity, and high-dynamic-range atomic magnetometer.
A detailed study of the Cs based magnetometer will be the subject of future investigation.

We have demonstrated, for the first time, a cesium-based laser–atomic oscillator (LAO).
Although the current clock-signal stability does not yet reach state-of-the-art performance, it represents an important step toward realizing the smallest active atomic clock. 
All the experimental results presented in this work were with a general-purpose commercial gain chip. 
Following our experimental demonstration of a Cs-LAO, we have been developing a custom gain chip both internally at Sandia National Laboratories and externally with a commercial partner, as further miniaturization is currently limited by the lack of a shorter gain chip.
We hope to realize a custom gain chip with the desired operating parameters in the near future. 
The fourth-order and seventh-order cavities used here rely on relatively large components, which introduce local temperature modulation and gradients, and consequent cavity-mode instabilities.
As a result, we obtain a short-term instability of $\sim$ 10$^{-10}$ and observe long-term thermal modulation of the clock signal on timescales ranging from tens of seconds to longer durations.
Implementing a first-order cavity should reduce both mechanical and thermal sensitivities.
Additionally, the LAO can function as an extremely sensitive magnetometer within the same clock architecture.
By locking the cavity to different pairs of Zeeman sublevels, we demonstrated magnetometer operation with a sensitivity of $\sim$ 100 fT/$\sqrt{\rm{Hz}}$ at 60\,Hz.
We expect the final LAO-based active atomic clock to be smaller than 1 cm$^3$.

\section*{AUTHOR DECLARATIONS}
\subsection*{Conflict of Interest}
The authors have no conflicts to disclose.
\subsection*{Author Contributions}
Y. Y. Jau conceived the experiments.  All authors contributed to the experiment building and carried them out.
S. Pandey and Y. Y. Jau wrote the manuscript. All authors reviewed the manuscript.

\begin{acknowledgments}
We are grateful to the Sterling McBride at SRI International for the research and development on the fabrication of high-pressure buffer gas vapor cells.
We acknowledged funding support from DARPA H6 program.
Sandia National Laboratories is a multimission laboratory managed and operated by National Technology \& Engineering Solutions of Sandia, LLC, a wholly owned subsidiary of Honeywell International Inc., for the U.S. Department of Energy’s National Nuclear Security Administration under contract DE-NA0003525.
The views, opinions and/or findings expressed are those of the author and should not be interpreted as representing the official views or policies of DARPA or the U.S. Government.
Distribution Statement "A" (Approved for Public Release, Distribution Unlimited).
\end{acknowledgments}

\section*{DATA AVAILABILITY}
The data that support the findings of this study can be made available from the corresponding author after approval from the relevant authorities.



\section*{REFERENCES}
\nocite{*}
\bibliography{CsLAO}

\end{document}